\documentclass[twocolumn,showpacs,preprintnumbers,amsmath,amssymb]{revtex4}
\usepackage{epsfig,graphicx,psfrag,here}
\newcommand{\bda}{\begin{\displaymath}\begin{array}{rl}}
\newcommand{\eda}{\end{array}\end{displaymath}}
\newcommand{\be}{\begin{equation}}
\newcommand{\ee}{\end{equation}}
\newcommand{\bdm}{\begin{displaymath}}
\newcommand{\edm}{\end{displaymath}}
\newcommand{\bea}{\begin{eqnarray}}
\newcommand{\eea}{\end{eqnarray}}
\newcommand{\no}{\nonumber \\}
\newcommand{\fs}{\; \; .}
\newcommand{\co}{\; \; ,}

\newcommand{\al}{&\!\!\!}
\begin{document}

\title{Mass and width of the lowest resonance in QCD}

\author{I.~Caprini}\affiliation{ National Institute of Physics and Nuclear
  Engineering, Bucharest, R-077125 Romania} 
\author{G.~Colangelo and H.~Leutwyler}
\affiliation{Institute for Theoretical Physics, University of 
Bern, Sidlerstr. 5, CH-3012 Bern, Switzerland}

\begin{abstract}
  We demonstrate that near the threshold, the $\pi\pi$ scattering amplitude
  contains a pole with the quantum numbers of the vacuum -- commonly referred
  to as the $\sigma$ -- and determine its mass and width within small
  uncertainties. Our derivation does not involve models or parametrizations,
  but relies on a straightforward calculation based on the Roy equation for
  the isoscalar $S$-wave.
\end{abstract}

\pacs{11.30.Rd, 11.55.Fv, 11.80.Et, 12.39.Fe, 13.75.Lb}
\maketitle

\thispagestyle{empty}

According to the Particle Data Group, the lowest resonance in the spectrum of
QCD carries angular momentum $\ell=0$ and isospin $I=0$. The state is listed
as $f_0(600)$ and is usually called the $\sigma$.  It manifests itself as a
pole on the second sheet of the isoscalar $S$-wave of $\pi\pi$ scattering.  We
denote this partial wave amplitude by $t_0^0(s)$.  The numbers for the pole
position found in the literature cover a very broad range. For recent
reviews, we refer to \cite{Ochs,Pennington,Bugg}. In fact, since such
a state is not easily accommodated in the multiplets expected for $q\bar{q}$
bound states and glueballs, some authors question its existence.

All of the pole determinations we are aware of rely on models and, moreover,
use specific parametrizations to perform the analytic continuation. In the
present paper, we instead rely on an equation which has been shown to follow
from first principles, the dispersive representation of the partial wave
amplitude $t^0_0(s)$ due to Roy \cite{Roy}: 
\bea\label{eqRoyt00}
t_0^0(s)\al=\al a+(s-4M_\pi^2) \,b 
+\int_{4M_\pi^2}^{\Lambda^2}\!\! ds' \left\{
  K_0(s,s')\,\mbox{Im}\,t_0^0(s')\right. \no\al\al\hspace{-3em}\left.+\,
K_1(s,s')\,\mbox{Im}\,t_1^1(s')+\,
K_2(s,s')\,\mbox{Im}\,t_0^2(s')\right\} + 
d_0^0(s)\,.\eea 
It amounts to a twice subtracted dispersion relation. Crossing symmetry
implies that both subtraction constants can be expressed in terms of the
$S$-wave scattering lengths: $a=a_0^0$, 
$b=(2\,a_0^0-5\,a_0^2)/(12 M_\pi^2)$. The integral describes the
curvature generated by the $S$- and $P$-waves below $\Lambda$ and the
so-called driving term $d_0^0(s)$ collects the dispersion integrals over the
higher partial waves ($\ell\geq2$), as well as the high energy end of the
integral over the $S$- and $P$-waves.

Similar equations hold for all other partial waves. Those for the $S$- and
$P$-waves amount to a set of coupled integral equations, which strongly
constrain the low energy properties of these waves \cite{BFP,ACGL,CGL}.
Previous work on the Roy equations concerned the behaviour on the real axis.
In Ref.~\cite{CGL}, a crude estimate of the mass and width of the $\sigma$ was
given, but as emphasized in several reviews (see e.g.~\cite{Ochs,Pennington}),
this estimate relies on a parametric extrapolation off the real axis and is
thus subject to a sizeable systematic uncertainty.

The present paper closes this gap. We show that the domain of validity of the
Roy equations can be extended to complex values of $s$ and use this extension
to (a) prove the existence of a second sheet pole close to the threshold and
(b) determine the position of this pole within rather small uncertainties. For
this purpose, we only need the particular equation quoted above. The explicit
expression for the kernels occurring therein reads: 
\bea\label{eqKernels}
K_0(s,s')&=&\frac{1}{\pi(s'\!-\!s)}+\frac{2\,L}{3\pi(s\!-\!4M_\pi^2)}-
\frac{5s'\!+\!2s\!-\!16M_\pi^2}{3\pi s'(s'\!-\!4M_\pi^2)}\,,\no K_1(s,s')&=&
\frac{6(s'\!+\!2s\!-\!4M_\pi^2)\,L}{\pi(s'\!-\!4M_\pi^2)(s\!-\!4M_\pi^2)}-
\frac{3(2s'\!+\! 3 s\!-\!4M_\pi^2)}{\pi s'(s'\!-\!4M_\pi^2)}\,,\no
K_2(s,s')&=&\frac{10\,L}{3\pi(s\!-\!4M_\pi^2)}-\frac{5(2s'\!-\!s\!-\!4M_\pi^2)}
{3\pi s'(s'\!-\!4M_\pi^2)}\,,\\
&&\hspace{-5.8em}\mbox{where}\hspace{1em} L=\ln\left(
  \frac{s+s'-4M_\pi^2}{s'}\right)\fs\nonumber\eea 
The first term in $K_0(s,s')$ accounts for the contributions generated by the
right hand cut.  The remainder of $K_0(s,s')$ describes those contributions
from the left hand cut that are also due to $\mbox{Im}\,t_0^0$, while those
from $\mbox{Im}\,t_1^1$ ($P$-wave) and $\mbox{Im}\,t_0^2$ (exotic $S$-wave) are
booked separately. 

For our analysis, it is essential that the dispersion integral is dominated
by the contributions from the low energy region: because the Roy equations
involve two subtractions, the kernels $K_n(s,s')$ fall off in proportion to
$1/s'^{\,3}$ for large $s'$. Note that the contributions from the
left hand cut play an important role here: Dropping these, 
$K_0(s,s')$ reduces to the first term, which falls off only with the first
power of $s'$. Taken by itself, the contribution from the right hand cut is
therefore sensitive to the poorly known high energy behaviour of
$\mbox{Im}\,t_0^0(s')$, but taken together with the one from the left hand
cut, the high energy tails cancel. 

Since the decomposition into partial waves is useful only at low energies, the
dispersion integral over the $S$- and $P$-waves in Eq.~(\ref{eqRoyt00}) is cut
off, the high energy tail being included in $d^0_0(s)$. As discussed
in Ref.~\cite{ACGL}, $d^0_0(s)$ is dominated by the contribution from the
\begin{figure}[thb]
\vspace{-2.3em}
\hspace{-3em}
\includegraphics[width=7.35cm,angle=-90]{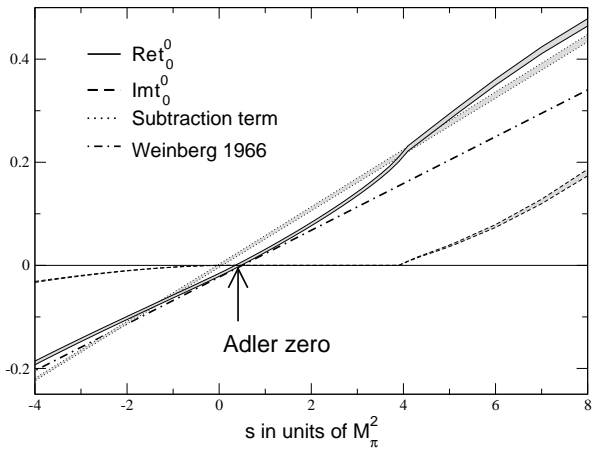}
\caption{\label{figsubthreshold}Behaviour of the amplitude near
  threshold. }   
\end{figure}
$D$-wave resonance $f_2(1270)$. Using the narrow width approximation,
expanding the relevant kernel in inverse powers of $s'=M_{f_2}^2$ and
retaining only the leading term of order $1/s'^{\,3}$, we obtain
\bea\label{eqdrivingterm}d_0^{\hspace{0.05em}0}(s)\simeq 
\frac{5(s-4M_\pi^2)(11 s+4 M_\pi^2)\,
\Gamma_{f_2\rightarrow\pi\pi}}{9 M_{f_2}^4\sqrt{\rule{0em}{1em}
    M_{f_2}^2-4M_\pi^2}} \fs \eea 
The experimental values for mass and width are $M_{f_2}=1275.4\pm 1.2$ MeV,
$\Gamma_{f_2\rightarrow\pi\pi}= 158.5\pm4.4$ MeV \cite{PDG 2004}.

We have performed a detailed evaluation of the driving term, which exploits
the Roy equations for the $D$-, $F$- and $G$-waves and accounts for the
contributions from the high energy tail. The calculation shows that, in the
low energy region we are considering in the present paper, the contributions
from partial waves with $\ell>2$ as well as those from energies above 1.4 GeV
are negligibly small and the narrow width approximation works: 
for $\Lambda\geq1.4$ GeV, the above simple formula is within the
uncertainties attached to the full result, which affect the outcome for the
pole position only by 1 or 2 MeV.

The most important feature in the low energy region is the occurrence of an
Adler zero. To leading order of chiral perturbation theory, the amplitude is
given by Weinberg's formula of 1966 \cite{Weinberg 1966},
\be\label{eqWeinberg}
t_0^0(s) =(2s-M_\pi^2)/(32\pi F_\pi^2)\fs\ee
In this approximation, the zero occurs at $s=\frac{1}{2}M_\pi^2$.  
Fig.\ref{figsubthreshold} shows the behaviour of the amplitude -- calculated
from Eq.~(\ref{eqRoyt00}) -- near the threshold and explains why the
theoretical predictions are so precise in this region: by far the most
important contribution stems from the subtraction term, which is linear in $s$
and is fixed by the two scattering lengths $a_0^0$, $a_0^2$.  Accordingly, the
values of $a_0^0$ and $a_0^2$ represent the most important ingredient of our
calculation. Since theory predicts these very accurately \cite{CGL}, 
\begin{figure}[thb]
\vspace{-2em}
\centering
\includegraphics[width=6.2cm,angle=-90]{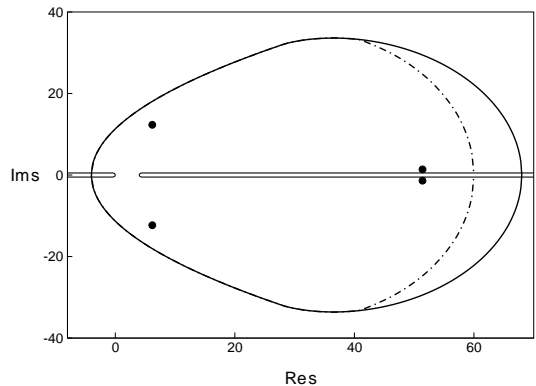}
\caption{\label{figdomain}Domain of validity of the Roy equations.}  
\end{figure}
\bea\label{eqprediction}
a_0^0=0.220\pm0.005\co\hspace{1em}a_0^2=-0.0444\pm0.0010 \co \eea 
the uncertainties in the subtraction term are very small. The main experiments
concerning the scattering lengths \cite{E865,DIRAC,NA48} are all in good
agreement with these predictions, but provide a stringent test only for the
first one. On the other hand, as will be discussed below, the experimental
information does suffice to evaluate the dispersion integrals in
Eq.(\ref{eqRoyt00}) to good accuracy.

The Roy equations represent the partial wave projections of the fixed-$t$
dispersion relations obeyed by the $\pi\pi$ scattering amplitude. These
relations are valid if the variable $t$ is contained in a Lehmann-Martin
ellipse with foci at $t=0$ and $t=4\,M_\pi^2-s'$ and right extremity at
$t=r(s')$, where $s'\geq 4\,M_\pi^2$ is the variable of integration. In the
Mandelstam representation, the size of the ellipse is limited by the
singularities due to the double spectral function. For $4\,M_\pi^2\leq s'\leq
20\,M_\pi^2$, the corresponding expression for $r(s')$ reads $r(s') = 16\,
s'/(s'-4\,M_\pi^2)$, while for $s'>20\,M_\pi^2$, the right extremity is at
$r(s')=4\,s'/(s'-16\, M_\pi^2)$.

The partial wave projection involves the values of the amplitude on the
interval $t=\frac{1}{2}\,(4\,M_\pi^2-s)(1-z)$, $0\leq z \leq 1$. For the Roy
equations to be valid at the point $s$, the corresponding interval must be
contained in the intersection of the ellipses characterized above. The
boundary of the domain $G$ where this is the case is shown as a solid line in
fig.~\ref{figdomain}, using pion mass units. The bound for $r(s')$ established
on the basis of axiomatic field theory \cite{Martin book,Mahoux} is only
slightly weaker. The dash-dotted line indicates the corresponding domain of
validity of the Roy equations. On the real axis, this domain reduces to the
range $-4\,M_\pi^2<s<60\,M_\pi^2$ obtained in Ref.~\cite{Roy}.

The values of the $S$-matrix element \bea
S^0_0(s)=1-2\sqrt{4\,M_\pi^2/s-1}\;t_0^0(s)\eea on the second sheet can be
calculated from those on the first sheet: unitarity implies the relation
\cite{Unitarity} \bea S_0^0(s)^{II}=1/S_0^0(s)^I\fs\eea The Roy equation
thus automatically also specifies the function $S_0^0(s)$ on the second
sheet, in the same domain of the $s$-plane \cite{bachir}.  In particular,
the amplitude contains a pole on the second sheet if and only if $S_0^0(s)$
has a zero on the physical sheet. So, all we need to do is numerically
evaluate Eq.~(\ref{eqRoyt00}) for complex values of $s$ in the domain $G$
where it has been shown to hold and find out whether or not $S_0^0(s)$ has
zeros there -- standard routines that solve equations numerically
immediately provide the answer.

As discussed in Ref.~\cite{CGL}, the Roy equations for the $S$ and
$P$-waves provide very firm control over the low energy behaviour of the
imaginary parts occurring in Eq.(\ref{eqRoyt00}). The numbers given in the
following are based on a new analysis of these equations which will be
described elsewhere. We could just as well have used the results in
Ref.~\cite{CGL} -- the numbers would barely change. For our central
representation of the scattering amplitude, the function $S_0^0(s)$ has two
pairs of zeros in $G$, one at $s = (6.2 \pm i \,12.3)\,M_\pi^2$, the other at
$s=(51.4\pm i\, 1.4)\,M_\pi^2$.  These are indicated in fig.~\ref{figdomain},
which may also be viewed as a picture of the second sheet -- the dots then
represent poles rather than zeros. In the following, we work with the
complex mass $m_\sigma\equiv
M_\sigma-\frac{i}{2}\,\Gamma_\sigma$, defined as the value of $\sqrt{s}$ at
the pole on the lower half of the second sheet. For the central solution, the
pole occurs at $m_\sigma=441- i\,272$ MeV, not far from the place where the
$\sigma$ was resurrected ten years ago \cite{Tornqvist and Roos}.

The second pole represents the
well-established resonance $f_0(980)$. In order to study the behaviour of the
amplitude there, we have extended the analysis of
Refs.~\cite{ACGL,CGL} to higher
energies. The BES data on the decay $J/\psi\rightarrow\phi\pi\pi$ \cite{BES
  f0(980)} have clarified the structure in this region. In this reference, the
sharp drop in the elasticity $\eta0_0(s)$ at the 
$K\bar{K}$ threshold is parametrized by means of the so-called Flatt\'{e}
formula, which describes the interference between the $K\bar{K}$ cut and the
pole from the $f_0(980)$. Given the elasticity, the Roy equations determine the
phase shift. It turns out that the solution closely follows the input: the
phase shift differs from the phase of the Flatt\'{e} formula only by a slowly
varying background. It does not come as a surprise, therefore, that our
calculation confirms the position of the pole in \cite{BES f0(980)} and
we do not discuss this further.

Finally, we estimate the uncertainty to be attached to the result obtained
from our central representation of the scattering amplitude. The Roy equations
imply that if the two subtraction constants $a_0^0$ and $a_0^2$ are treated as
known, the low energy properties of the isoscalar $S$-wave depend almost
exclusively on a single parameter, which may be identified with the value of
the phase at 800 MeV and which we denote by
$\delta_A\equiv\delta_0^0(800\,\mbox{MeV})$. Accordingly, it is convenient to
represent the position of the pole as
\bea\label{eqmsigma} m_\sigma \al=\al m_0+m_1\,\Delta a_0^0+m_2\,\Delta
a_0^2+m_3\, \Delta \delta_A \,,\\
\Delta a_0^0\al=\al(a_0^0-0.22)/0.005\,,\;\Delta a_0^2=(a_0^2+0.0444)/0.001\,,
\no \Delta \delta_A\al=\al(\delta_A-82.3)/3.4\,.  \nonumber\eea 
The term $m_0$ represents the value of the complex mass obtained if the
scattering lengths are fixed at the central values of the prediction
(\ref{eqprediction}), while the phase at 800 MeV is set equal to the central
value obtained from the phenomenology of the phase difference between
$\delta_1^1$ and $\delta_0^0$ \cite{ACGL}.  The coefficients $m_1$, $m_2$ and
$m_3$ describe the sensitivity of the result to a change in these variables --
for the small changes of interest, the linear approximation is adequate.

The residual noise in $m_0$ is small, for the following reasons. (a) The
integral over $\mbox{Im}\,t^0_0$ is dominated by the region below the
$K\bar{K}$ threshold. If $a_0^0$, $a_0^2$ and $\delta_A$ are held fixed, the
Roy equations constrain the integrand very strongly there. (b) The
contribution from the $P$-wave is dominated by the $\rho$. Since the
experimental information about this state is excellent, the integral over
$\mbox{Im}\, \,t_1^1$ is known very well. (c) In $\mbox{Im}\,t^2_0$, the
uncertainties are larger, but the entire contribution from this wave is small.
(d) As stated above, the uncertainties in the higher partial waves and in the
contributions from the high energy tail of the dispersion integrals affect the
result for $m_\sigma$ only at the level of 1 or 2 MeV. In our opinion, the
estimate $m_0=(441\pm 4)-i\,(272\pm 6)$ MeV generously covers the
uncertainties. For the other coefficients in Eq.~(\ref{eqmsigma}),
we obtain $m_1=-2.4+i\,3.8$, $m_2=0.8-i\,4.0$, $m_3 = 5.3+i\,3.3$ (values in
MeV).

The result for the pole position does not change significantly if the
imaginary parts are evaluated with a phenomenological representation of the
data. We illustrate this with the parametrization of the scattering
amplitude proposed in appendix A of Ref.~\cite{PY}. Evaluating the
dispersion integrals over $\mbox{Im}\,t_0^0$, $\mbox{Im}\,t_1^1$ and
$\mbox{Im}\,t_0^2$ with the central parametrization of that reference, we
find that the pole occurs at $445-i\,241$ MeV. Although this representation
differs significantly from our central solution, the result for the pole
position agrees with formula (\ref{eqmsigma}): for the values $a_0^0=0.23$,
$a_0^2=-0.048$ and $\delta_A=90.9^\circ$ that correspond to this
parametrization, the formula yields $m_\sigma=447 -i\,242$ MeV.

This confirms that the position of the pole from the $\sigma$ is indeed
controlled by three observables. Only one of these, $a_0^0$, is known
experimentally to good precision. For the second one, $a_0^2$, the sharp
theoretical prediction in Eq.~(\ref{eqprediction}) was recently confirmed by
an evaluation on the lattice: $a^2_0 = -0.0426(6)(3)(18)$, where the three
brackets give the statistical, systematic and theoretical errors, respectively
\cite{Beane} (cf. also \cite{Yamazaki:2004qb}). To stay on the conservative
side with the third parameter, we use $\delta_A=82.3^\circ\stackrel{+
  10^\circ}{{}_{-4^\circ}}$.  Compared to the uncertainty from this source,
the noise in the term $m_0$ is negligible.  With the theoretical prediction
for the scattering lengths, formula (\ref{eqmsigma}) then leads to our final
result
\bea\label{eqfinal}
M_\sigma =441 \stackrel{+16}{_{-8}}\,\mbox{MeV}\,,\;\; \Gamma_\sigma=544
\stackrel{+18}{_{-25}}\,\mbox{MeV}\fs\eea 
We conclude that the same theoretical framework that leads to incredibly sharp
predictions for the threshold parameters of $\pi\pi$ scattering \cite{CGL}
also requires the occurrence of a pole on the lower half of the second sheet,
with the quantum numbers of the vacuum, not much above the threshold, but
quite far from the real axis: the width of the $\sigma$ is larger than the
width of the $\rho$ by a factor of 3.7. The parametric extrapolation used in
Ref.~\cite{CGL} led to somewhat higher values, for the mass as well as for the
width -- the difference amounts to about 1 standard deviation. The uncertainty
in the result (\ref{eqfinal}) stems almost exclusively from $\delta_A$. The
range adopted for this observable covers all of the phenomenological
parametrizations we are aware of. (Energy-independent analyses have a broader
scatter of values, but even the most extreme \cite{Kaminski:2001hv} is less
than $2\,\sigma$ away from our central value.) It can be reduced substantially
if the data that underly these parametrizations are compared with the
solutions of the Roy equations. We intend to describe this elsewhere.

The experimental information concerning $a_0^2$ is meagre, but since the real
part of the coefficient $m_2$ is very small, this does not affect the value of
$M_\sigma$: as far as the real part of the pole position is concerned, the
result remains practically unchanged if the theoretical predictions for the
scattering lengths are replaced by the experimental constraints on these
\cite{E865,DIRAC,NA48}. 

Many of the determinations of the mass and width of the $\sigma$ listed by
the Particle Data Group neglect the left hand cut. In the language of
Eq.~(\ref{eqRoyt00}), this approximation amounts to replacing the kernel
$K_0$ by the first term in Eq.~(\ref{eqKernels}) and dropping $K_1$, $K_2$
as well as $d_0^0$. For definiteness, we fix the subtraction constants $a$
and $b$ such that the ``exact'' and approximate representations agree at
the threshold and at $\sqrt{s_A}= 800\,\mbox{MeV}$. At the energies of
interest, the difference is then well described by the parabola
$t_0^0(s)_{\mbox{lhc}}\simeq c\,(s-4\,M_\pi^2)(s_A-s)$, with $c=
0.5\,\mbox{GeV}^{-4}$. Removing this term from $t_0^0(s)$ has a rather
drastic effect: the pole then occurs at $m_\sigma \simeq 500 - i\, 260$
MeV. The amplitude obtained in this manner cannot be taken at face value,
because it violates unitarity: dropping the left hand cut necessarily also
distorts the imaginary part. We do not pursue this further.  The above
expression for the corresponding curvature shows that the left hand cut
cannot be neglected -- the pole is not sufficiently far away from it for
this approximation to be meaningful.

It is difficult to understand the properties of the lowest resonances in terms
of the degrees of freedom of the quarks and gluons. The physics of the
$\sigma$ is governed by the dynamics of the Goldstone bosons: The properties
of the interaction among two pions are relevant \cite{Markushin and Locher}. A
qualitative explanation for 
the occurrence of the $\sigma$ was given in Ref.~\cite{CGL}, on the basis of
current algebra, spontaneous symmetry breakdown and unitarity. The properties
of the resonance $f_0(980)$ are also governed by Goldstone boson dynamics --
two kaons in that case. It would be of considerable interest to apply the
above analysis to the Roy-Steiner equation \cite{Buettiker Descotes
  Moussallam} for the $K\pi$ $S$-wave with $I=\frac{1}{2}$. This should
clarify the situation with the $\kappa$.

\acknowledgments We are indebted to David Bugg for numerous comments
concerning the issues discussed above, in particular for providing us with
detailed information about the structure of the amplitude near the $K\bar{K}$
threshold. Also, we wish to thank Peter Minkowski and Wolfgang Ochs for
informative discussions. This work is supported by Schweizerischer
Nationalfonds, by the Romanian MEdC under Contract CEEX 05-D11-49 and by the
EU ``Euridice'' program under code HPRN-CT2002-00311.


\begin{thebibliography}{99}

\bibitem{Ochs}W.~Ochs, in {\it Hadron Spectroscopy: Tenth International
    Conference}, edited by E.~Klempt, H.~Koch and H. Orth, 
  AIP Conf.\ Proc.\ No.\ 717 (AIP, New York, 2004), p.~295.

\bibitem{Pennington}M.~R.~Pennington,
  arXiv:hep-ph/0509265.

\bibitem{Bugg}D.~V.~Bugg, in {\it Hadron Spectroscopy: Eleventh International
  Conference on Hadron Spectroscopy}, edited by A.~Reis, C.~G\"obel, J.~de
  S\'a Borges and J.~Magnin, AIP Conf.~Proc.~No.~814 (AIP, New York, 2006),
  p.~78.

\bibitem{Roy}
S.~M.~Roy,
  Phys.\ Lett.\ B {\bf 36} (1971) 353.

\bibitem{BFP}J.~L.~Basdevant, C.~D.~Froggatt and J.~L.~Petersen,
Nucl.\ Phys.~{\bf B 72} (1974) 413.

\bibitem{ACGL}
B.~Ananthanarayan, G.~Colangelo, J.~Gasser and H.~Leutwyler,
  Phys.\ Rept.\  {\bf 353} (2001) 207.

\bibitem{CGL} 
  G.~Colangelo, J.~Gasser and H.~Leutwyler,
  Nucl.\ Phys.\ B {\bf 603} (2001) 125.

\bibitem{PDG 2004}
 S.~Eidelman {\it et al.}  [Particle Data Group],
  Phys.\ Lett.\ B {\bf 592} (2004) 1.

\bibitem{Weinberg 1966}
 S.~Weinberg,
  Phys.\ Rev.\ Lett.\  {\bf 17} (1966) 616.

\bibitem{E865}S.~Pislak {\it et al.}  [BNL-E865 Collaboration],
Phys.\ Rev.\ Lett.\  {\bf 87} (2001) 221801
Phys.\ Rev.\ D {\bf 67} (2003) 072004.

\bibitem{DIRAC}
 B.~Adeva {\it et al.}  [DIRAC Collaboration],
  Phys.\ Lett.\ B {\bf 619} (2005) 50.

\bibitem{NA48}
J.~R.~Batley {\it et al.}  [NA48/2 Collaboration],
Phys.~Lett.~B {\bf 633} (2006) 173.

\bibitem{Martin book}A.~Martin, {\it Scattering Theory: Unitarity,
Analyticity and Crossing}, Lecture Notes in Physics, Vol.~3, 
(Springer-Verlag, Berlin, 1969).

\bibitem{Mahoux}
G.~Mahoux, S.~M.~Roy and G.~Wanders,
  Nucl.\ Phys.\ B {\bf 70} (1974) 297.
    
\bibitem{Unitarity}
For a derivation, see for instance Ref.~\cite{BFP}

\bibitem{bachir}
We thank Bachir Moussallam for this remark.

\bibitem{Tornqvist and Roos}N.~A.~Tornqvist and M.~Roos,
  Phys.\ Rev.\ Lett.\  {\bf 76} (1996) 1575.

\bibitem{BES f0(980)}
M.~Ablikim {\it et al.}  [BES Collaboration],
Phys.\ Lett.\ B {\bf 607} (2005) 243.


\bibitem{PY} J.~R.~Pel\'aez and F.~J.~Yndur\'ain,
Phys.\ Rev.\ D {\bf 71} (2005) 074016.

\bibitem{Beane}S.~R.~Beane, P.~F.~Bedaque, K.~Orginos and M.~J.~Savage  
[NPLQCD Collaboration],
hep-lat/0506013.

\bibitem{Yamazaki:2004qb}
T.~Yamazaki {\it et al.}  [CP-PACS Collaboration],
Phys.\ Rev.\ D {\bf 70} (2004) 074513.

\bibitem{Kaminski:2001hv}
R.~Kaminski, L.~Lesniak and K.~Rybicki,
Eur.\ Phys.\ J.\ directC {\bf 4} (2002) 1.
  
\bibitem{Markushin and Locher}
V.~E.~Markushin and M.~P.~Locher, in {\it Workshop in Hadron Spectrocopy},
Frascati Physics Series, Vol.~XV (Laboratori Nazionali di Frascati, Rome,
1999), p.~229. 

\bibitem{Buettiker Descotes Moussallam} P.~Buettiker, S.~Descotes-Genon and
B.~Moussallam,
Eur.\ Phys.\ J.\ C {\bf 33} (2004) 409.

\end{thebibliography}
\end{document}